\documentstyle[11pt,newpasp,twoside,epsf]{article}
\def\edcomment#1{\iffalse\marginpar{\raggedright\sl#1\/}\else\relax\fi}
\reversemarginpar

\newcommand{\der}[2]  { \frac{{\rm d}#1}{{\rm d}#2} }

\newcommand{\dif}     {{\rm d}}

\begin{document}
\title{The Hydrodynamics and Chemical Evolution of Starburst-driven Outflows}
\author{Sergiy Silich}
\affil{Instituto Nacional de Astrof\'{\i}sica Optica y Electr{\'o}nica, 
Aptdo. 51 y 216, 72000 Puebla, Pue. M{\'e}xico}
\author{Guillermo Tenorio-Tagle}
\affil{Instituto Nacional de Astrof\'{\i}sica Optica y Electr{\'o}nica, 
Aptdo. 51 y 216, 72000 Puebla, Pue. M{\'e}xico}

\begin{abstract}
The hydrodynamics and intrinsic properties of galactic-scale 
gaseous outflows generated by violent starbursts are 
thoroughly discussed, taking into account the hot gas chemical 
evolution and  radiative cooling. We also discuss the observational 
properties of supergalactic winds in X-rays and visible line 
regimes, derived from the hydrodynamic calculations.

\end{abstract}

\section{Introduction}

Massive starbursts, well localized short episodes of violent star 
formation, are among the intrinsic characteristics of the evolution 
of galaxies. They are found both at high and intermediate  redshifts 
as well as in galaxies in the Local universe. 

The large kinetic luminosity produced by violent bursts of star 
formation is well known to drastically affect the surrounding
interstellar medium (ISM), generating giant superbubbles, supershells 
and in extreme cases supergalactic winds. Powerful gas outflows from
star forming regions are important redistributors of the ISM mass and 
momentum. On the other hand, the gas ejected by supernova explosions
contains the products of massive stars. 
Therefore gaseous outflows initiated by starburst affect also the 
chemical evolution of the galactic ISM and the intergalactic medium. 

Supergalactic winds are expected to be detectable thanks to their 
extended X-ray emission. Indeed, diffuse X-ray emission 
associated with starburst galaxies has been detected by ROSAT, ASCA,
BeppoSAX, Chandra and XMM missions in several starburst sources. 
However, the origin of the soft X-ray emission, its connection with 
H$_{\alpha}$ filaments and the chemical composition of the X-ray 
emitting gas, remain ambiguous.

The paper deals with three central aspects of the resultant outflows.
The chemical evolution expected in the interior of superbubbles.
The conditions required to establish a supergalactic wind 
and the intrinsic properties and observational manifestations of
fully developed supergalactic winds.

\section{Hot gas chemical composition}

The metallicity of the interstellar and intergalactic media reflects 
the history of star formation in the Universe. Galactic supershells 
and superwinds have been advocated as major mechanisms which 
re-distribute heavy elements throughout galactic discs and the 
intergalactic space. However, not until recently theoretical
models did not care about the hot bubble chemical evolution and 
associated the hot gas metallicity with the metallicity of the 
surrounding ISM. 

The enrichment caused by supernova products has been incorporated
in the models by Silich et al. (2001). The calculations
assume an instantaneous burst of star formation with a Salpeter 
initial mass function with an exponent $\alpha$ and incorporate 
the oxygen and iron
yields Y$_{O,Fe}$ derived from stellar evolution models. 
The mass of oxygen (and iron) released by the stellar cluster 
of total mass M$_{SB}$ with upper and lower cutoff masses 
M$_{up}$ and M$_{low}$,  during the evolutionary time t is
\begin{equation}
      \label{eq.1}
M_{ej}(O,Fe) = 
\frac{(\alpha -2) M_{SB}}{M_{low}^{2-\alpha} - M_{up}^{2-\alpha}}
\int_{M_{\star}(t)}^{M_{up}} Y_{O,Fe}(m) m^{-\alpha} {\rm d}m.
\end{equation}
The ejected metals were assumed to be effectively mixed with the
stellar hydrogen envelopes and matter evaporated from the shock
driven outer shell. The mean hot gas metallicity measured in Solar 
units is then
\begin{equation}
      \label{eq.2}
Z_{O,Fe} = \frac{M_{ej}(O,Fe)/Z_{\odot}(O,Fe) + Z_{ISM} M_{ev}}
                        {M_{ev} + M_{ej}},
\end{equation}
where Z$_{\odot}$(O,Fe) are the oxygen and iron solar content,
Z$_{ISM}$ is the interstellar gas metallicity, M$_{ej}$ and 
M$_{ev}$ are the total ejected mass and the mass loaded
by thermal evaporation. The results of the calculations
are presented in Figure 1.
\begin{figure}[htbp]
\plotfiddle{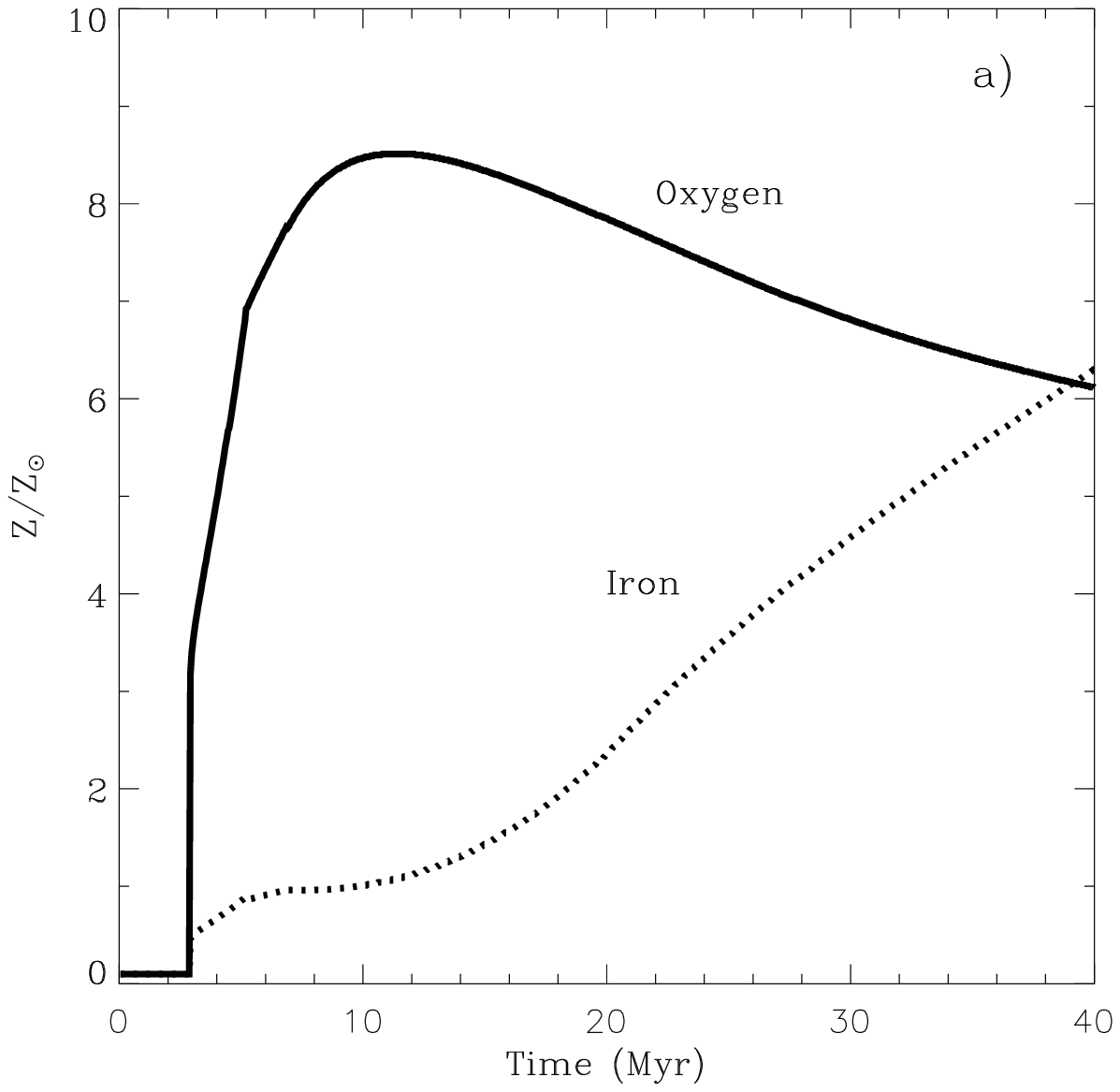}{2.75cm}{0}{50}{50}{-240}{-280}
\plotfiddle{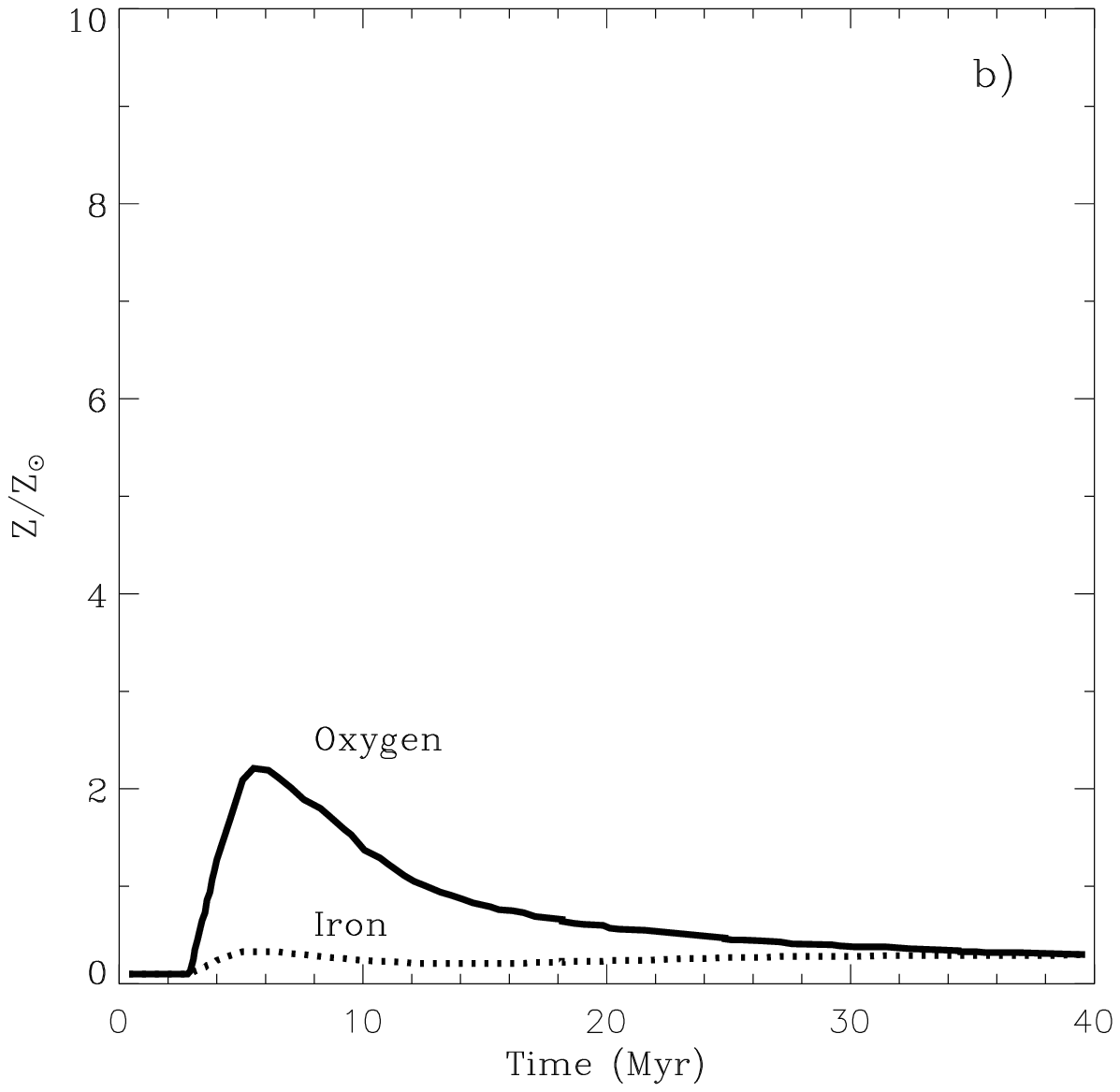}{2.75cm}{0}{50}{50}{-60}{-189}
\caption{The oxygen and iron contents of superbubbles as a function
         of time, in solar units. Panels a) and b) present the hot 
         gas metallicity without and when considering the evaporation
         of the cold outer shell, respectively.}
\end{figure}
Figure 1a displays the extreme case without any mass loading and
therefore represents the upper limit for the metallicity of a
superbubble. Figure 1b represents a more realistic case, with 
dilution caused by the low metallicity ISM. In this case the
oxygen abundance reaches its maximum Z$_O \sim 2$Z$_{\odot}$ 
after 6Myr of evolution and then falls slowly to the solar value 
after 10Myr to finally approach 0.1Z$_{\odot}$
at the end of the starburst activity.

These results are in contradiction with the previous estimates
of the X-ray emitting gas metallicities based on the analysis of the
ROSAT data. However, as mentioned by Strickland \& Stevens
(1998) the low, subsolar metallicities most probably come
from the single temperature model used to fit
the ROSAT spectra. Another possible solution to the low abundance
problem  has been indicated by Breitschwerdt (2003), who provided a 
detailed analysis of the non-equilibrium ionization model. 

The recent estimates of the hot gas metal abundances 
in the starburst galaxy NGC 1569 by Martin, Kobulnicky, \& Heckman
(2002), are based on two temperature model and show an excellent 
agreement with our theoretical predictions. The $\alpha$ element 
metallicities for a starburst with the age between 10Myr to 20Myr 
have been best fitted by Z$_{\alpha} = 1.0$Z$_{\odot}$, and abundances 
higher than solar have not been excluded. 
 
The ultimate fate of metals released within the starburst regions 
has been intensively debated in recent years (see Kunth; this volume
and references therein). It is highly dependent on the kinematic 
properties of the host galaxy (Silich \& Tenorio-Tagle 2001).
Supergalactic winds rapidly develop in fast rotating systems with 
a disc-like ISM distribution and are less likely in slow rotating 
galaxies with a thick-disc or a spherical gas distribution.
The metallicity-flattening relation revealed by Barazza \& Binggeli 
(2002) that shows that rounder dwarf ellipticals tend to be more 
metal-rich, may support these conclusions.

\section{Supergalactic wind}

Energetic starbursts, evolving in a flat disc-like ISM, are 
able to drive their associated shock waves to the outskirts of their 
host galaxies, leading to the development of supergalactic winds. 
It is usually assumed that within the region of star formation
the matter ejected by strong stellar winds and supernova explosions is 
fully thermalized by means of random collisions. This generates the 
high central temperature and large overpressure that initiates the
metal-rich gas outflow. There, the mean total energy L$_{SB}$ 
and mass \.M$_{SB}$ deposition rates control, together with the 
actual size of the star forming region R$_{SB}$, the properties of the 
resultant outflow. After crossing r = R$_{SB}$ the gas is immediately 
accelerated by steep pressure gradients and rapidly reaches 
its terminal velocity V$_t$. This is due to a fast 
conversion of thermal energy, into kinetic energy of the resultant 
wind. The free wind analytic solution of Chevalier \& Clegg (1985) 
assumes that the thermalized gas freely expands out of 
the star forming region and then adiabatically cools down approaching 
an r$^{-4/3}$ temperature distribution.

\subsection{The steady state solution}

Here we study the true physical properties of such well developed
free wind outflows, taking into consideration strong radiative cooling.
Following Chevalier \& Clegg (1985) we assume a spherically symmetric 
wind, unaffected by the gravitational pull caused by the central star 
cluster. The equations that govern the steady gas outflow away from  
the star forming region are:
\begin{eqnarray}
      \label{eq.3a}
      & & \hspace{-0.5cm}
\frac{1}{r^2} \der{}{r}\left(\rho u r^2\right) = 0 ,
      \\[0.2cm]
      \label{eq.3b}
      & & \hspace{-0.5cm}
\rho u \der{u}{r} = - \der{P}{r} ,
      \\[0.2cm]
     \label{eq.3c}
      & & \hspace{-0.5cm}
\frac{1}{r^2} \der{}{r}{\left[\rho u r^2 \left(\frac{u^2}{2} +
\frac{\gamma}{\gamma - 1} \frac{P}{\rho}\right)\right]} = - Q,
\end{eqnarray}
where r is the spherical radius, and $u(r), \rho(r)$ and $P(r)$ are 
the wind velocity, density and thermal pressure, respectively. Q is
the cooling rate ($Q = n^2 \Lambda$) where $n$ is the wind number 
density and $\Lambda$ is the cooling function.
\begin{figure}[h]
\plotfiddle{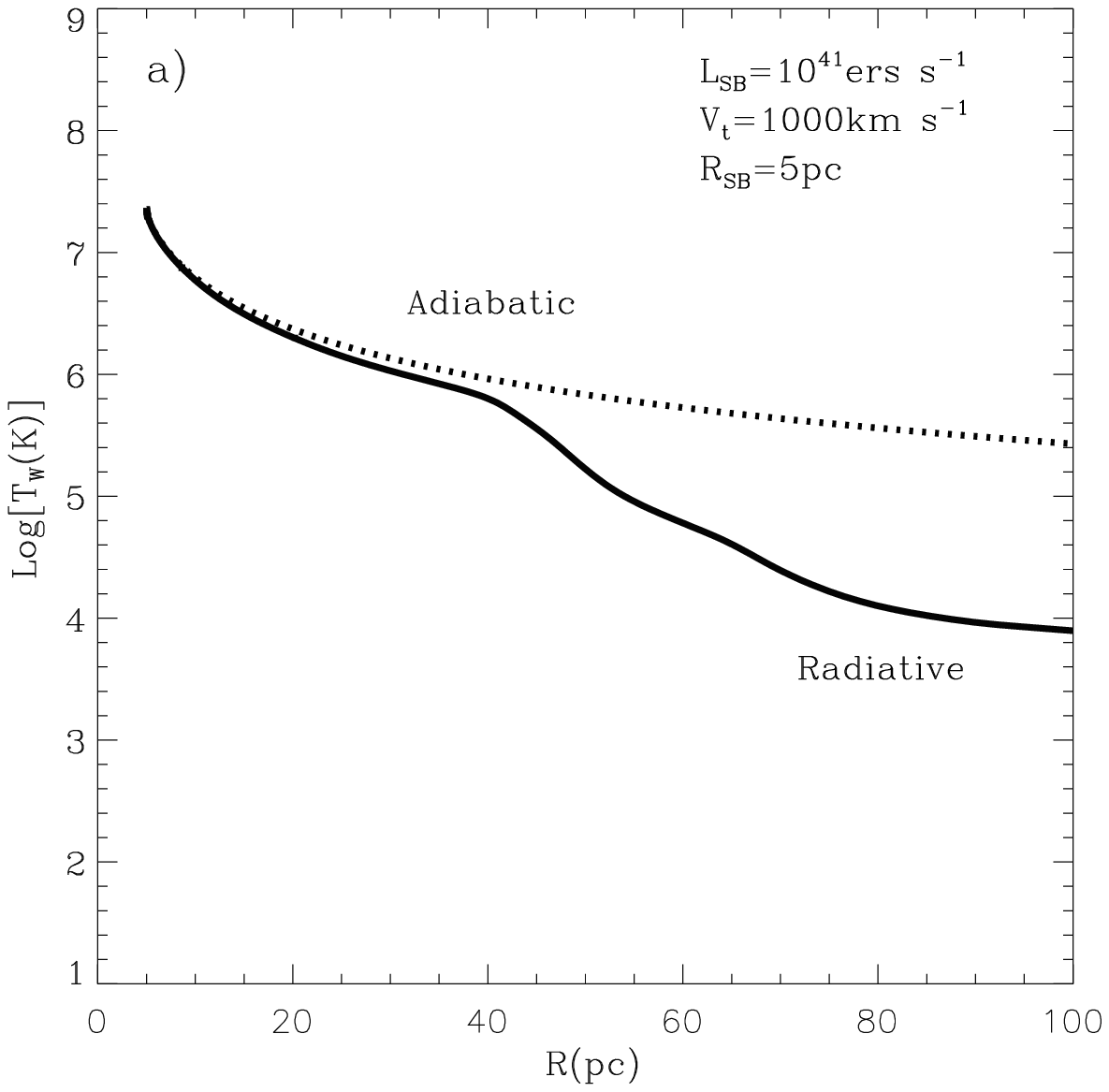}{2.75cm}{0}{50}{50}{-240}{-280}
\plotfiddle{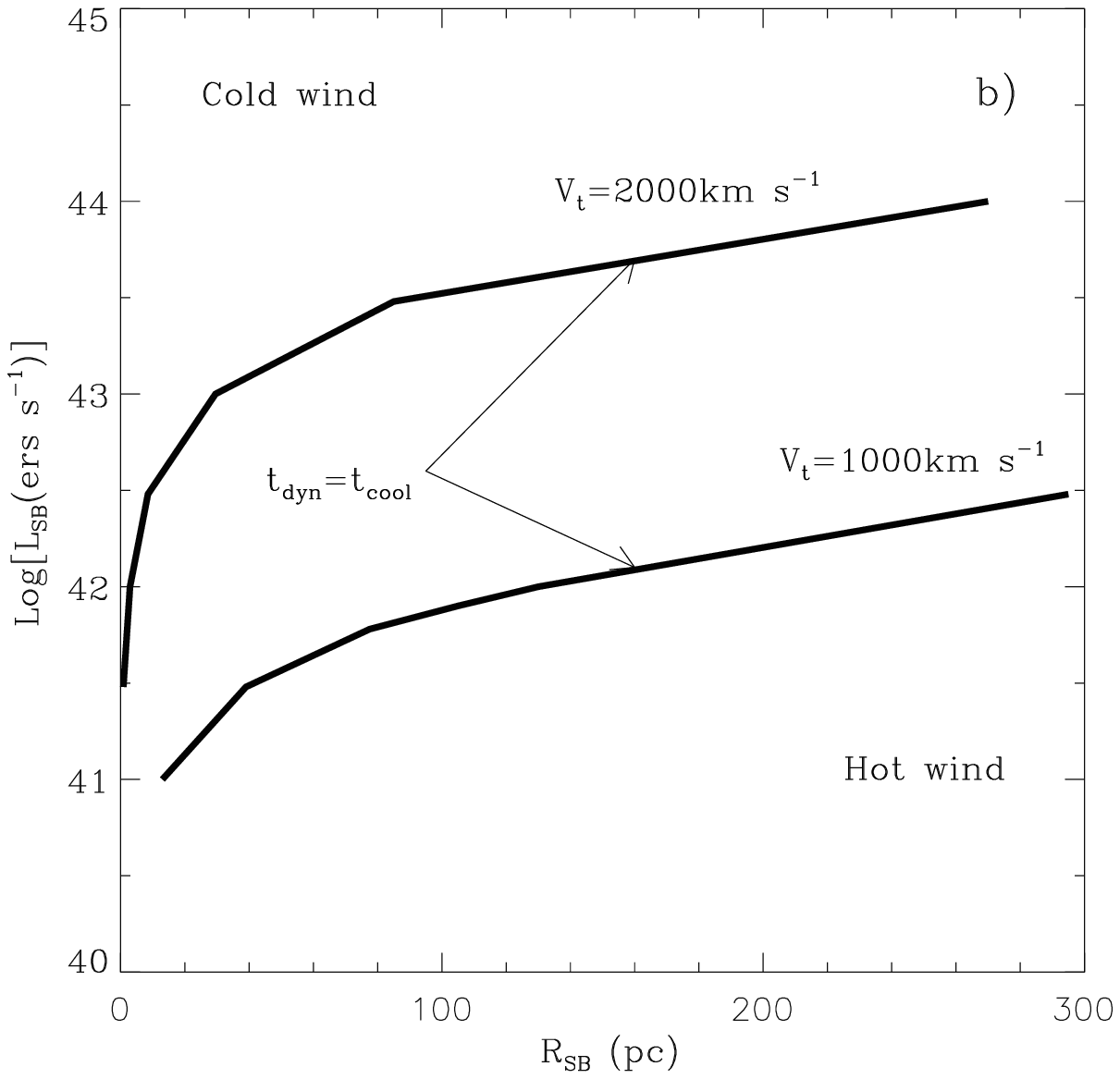}{2.75cm}{0}{50}{50}{-60}{-189}
\caption{Superwinds. a) Temperature distribution for adiabatic and
strongly radiative solutions. b) The two possible outcomes, hot
and cold winds, as function of the parameter space.}
\end{figure}

\subsection{The impact of radiative cooling}

A first order of magnitude estimate of whether or not radiative cooling 
could affect the thermodynamics of superwinds, results  from a 
comparison of the radiative cooling time with the characteristic dynamical 
time scale : 
\begin{equation}
      \label{eq.4} 
\tau_{cool}(r) = \frac{3 k T}{n \Lambda} , \qquad
\tau_{dyn}(r) = \int_{R_{SB}}^r \frac{\dif{r}}{u(r)} .
\end{equation} 

Our calculations show that both the velocity and the density distributions
remain practically unaffected by radiative cooling. However, 
if radiative cooling becomes efficient, the temperature distribution 
strongly deviates away from the adiabatic solution forcing the gas to 
soon reach temperature values of the order of 10$^4$ K (see Figure
2a).

Figure 2b implies that radiative  cooling may become efficient for 
compact massive star clusters. If the initial wind parameters 
(L$_{SB}$ and R$_{SB}$) intersect below the corresponding terminal 
velocity curve, cooling will be inefficient  and deviations from the 
adiabatic solution would be negligible. However, if the initial 
wind parameters intersect above the corresponding terminal velocity
curve, radiative cooling is expected to become important, causing a
major impact on the appearance of superwinds.  
\begin{figure}
\plotfiddle{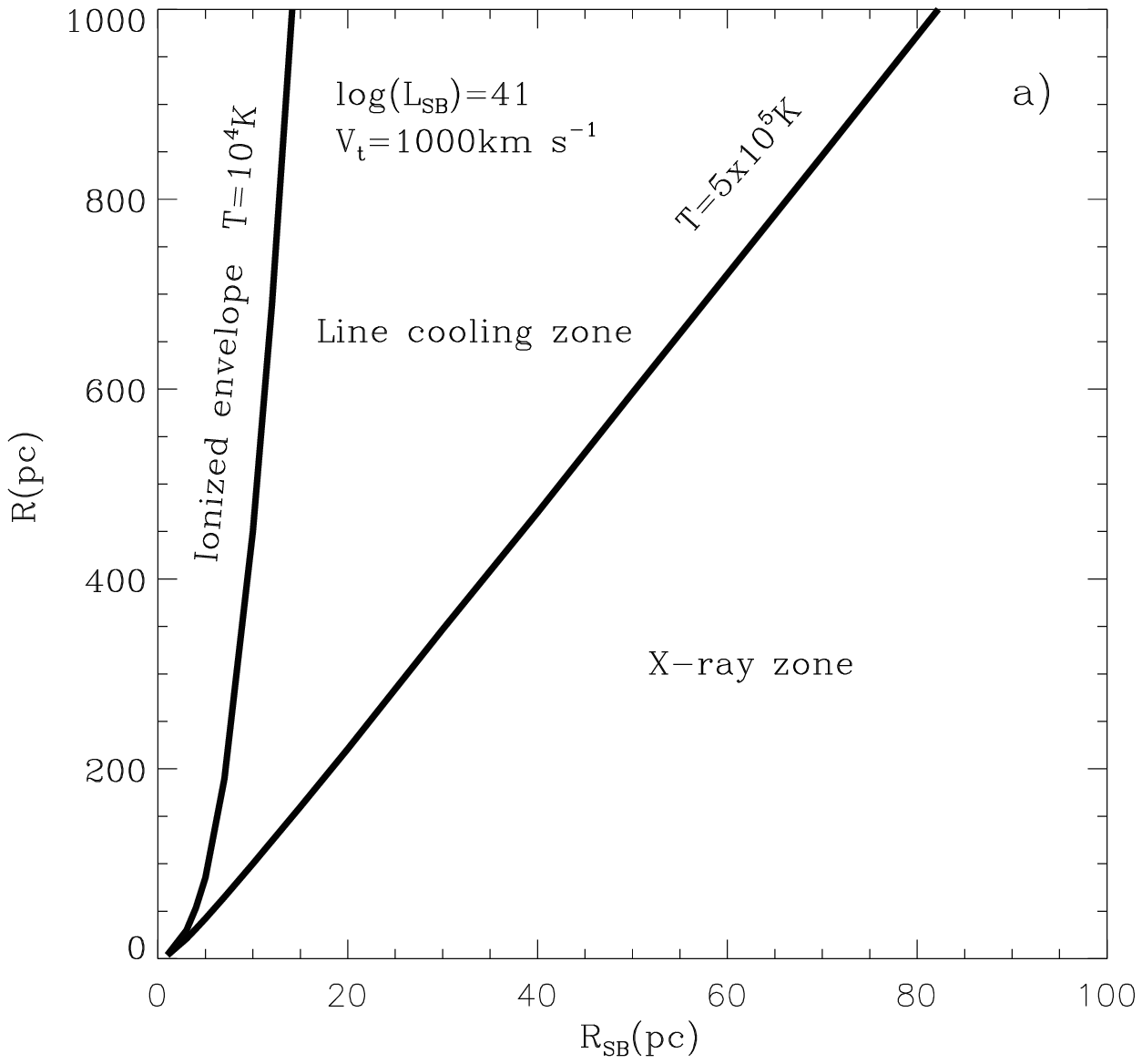}{2.75cm}{0}{50}{50}{-240}{-280}
\plotfiddle{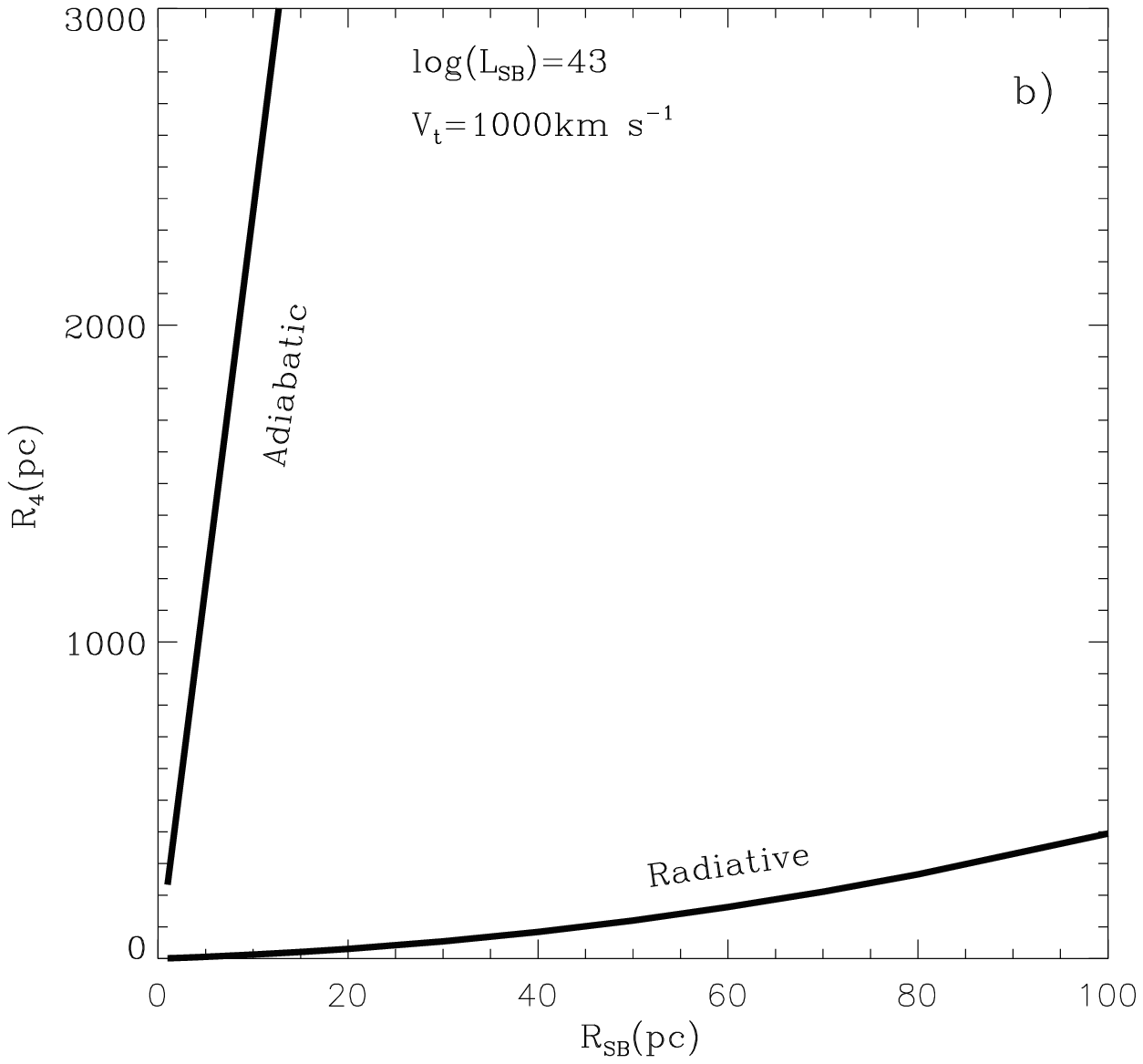}{2.75cm}{0}{50}{50}{-60}{-189}
\caption{Superwinds. a) Internal structure for low luminosity 
starbursts, as a function of the size of the star forming region.
b) The lines indicate the place where the gas acquires a T = $10^4$K
in adiabatic and radiative calculations of powerful starburst
(L$_{SB} = 10^{43}$ erg s$^{-1}$).}
\end{figure}

\subsection{The radiative mode observational implications}

Cooling provides significant modifications of a superwind internal 
structure (see Figure 3), bringing the outer boundary of the
X-ray emitting zone and inner boundary of the cold envelope
closer to the star cluster. This in turn 
modifies the predicted X-ray and H$_{\alpha}$ luminosities. 
The X-ray emission vanishes at large distances and arises
only from zones close to the star forming region, thus leading
to a drop in the total X-ray luminosity. 

The trend of the  H$_{\alpha}$
emission is different. When cooling sets in, much denser superwind 
layers acquire  temperatures below $10^4$K. These are to be
ionized by the central star cluster and by the soft X-ray 
photons to become visible in the optical line regime. This causes
an increase in the H$_{\alpha}$ luminosity. For compact star clusters 
the difference between the adiabatic and radiative model predictions
may reach almost two orders of magnitude.

\section{Conclusions}

The results from the calculations presented here imply that:

The metallicity of superbubbles vary with time and can easily exceed the 
solar value even if the host galaxy ISM has a low metal abundance. 
The effects of metal contamination are most noticeable during the first 
10 Myr of the bubble evolution.

Galactic superwinds driven by compact and powerful starbursts 
undergo catastrophic cooling close to the star 
cluster surface and establish a temperature distribution  
radically different to that predicted by adiabatic calculations. 

The fall of the superwind temperature leads to a smaller zone 
radiating in X-rays and decreases the superwind X-ray luminosity.

At the same time cooling brings the inner radius of the
warm ionized gas envelope closer to the star cluster.
This increases the estimated H$_{\alpha}$ luminosity and 
predicts a low-intensity broad ($\sim 1000$ km s$^{-1}$) 
line emission component.

\acknowledgments 

The authors feel honoured to have attended this special meeting and 
acknowledge financial support from CONACYT (M{\'e}xico) grant 36132-E.

\section*{\bf Discussion}

\noindent
{\it Casiana Mu\~noz-Tu\~n\'on:} Do your results imply that cooling
     may affect the evolution of superbubbles ? \\

\noindent
{\it Silich:} No, we have considered the late post-blowout stage, when
     the reverse shock reaches the outskirts of the galaxy and the 
     outflow approaches a free-wind steady state regime. \\

\end{document}